\documentclass[twocolumn,showpacs,preprintnumbers,amsmath,amssymb,footinbib]{revtex4-1}

\usepackage{graphicx}% Include figure files
\usepackage{epstopdf,mathrsfs}
\begin{document}

%\preprint{APS/123-QED}

\title{Conversion efficiency in the process of co-polarized spontaneous four-wave mixing} % Force line breaks with \\

\author{Karina Garay-Palmett$^{1,2}$, Alfred B. U'Ren$^{1}$, Ra\'ul Rangel-Rojo$^2$}
\affiliation{
$^{1}$Instituto de Ciencias Nucleares, Universidad Nacional Aut\'onoma de M\'exico, apdo. postal 70-543, M\'exico 04510 DF \\
$^{2}$Departamento de \'Optica, Centro de Investigaci\'on Cient\'{\i}fica y
de Educaci\'on Superior de Ensenada, Apartado Postal 2732, Ensenada,
BC 22860, M\'exico.}

\date{\today}% It is always \today, today,
% but any date may be explicitly specified

%\newcommand{\epsfg}[2]{\centerline{\scalebox{#2}{\epsfbox{#1}}}} %

\begin{abstract}
We study the process of co-polarized spontaneous four-wave mixing in
single-mode optical fibers, with an emphasis on an analysis of the
conversion efficiency.  We consider both the monochromatic-pumps and
pulsed-pumps regimes, as well as both the degenerate-pumps and
non-degenerate-pumps configurations.  We present analytical
expressions for the conversion efficiency, which are given in terms
of double integrals.  In the case of pulsed pumps we take these
expressions to closed analytical form with the help of certain
approximations.  We present results of numerical simulations, and
compare them to values obtained from our analytical expressions, for
the conversion efficiency as a function of several key experimental
parameters.
\end{abstract}

% PACS, the Physics and Astronomy
% Classification Scheme.
%\keywords{Photons nonclassical states, Quantum entanglement, Four-wave mixing}%Use showkeys class option if keyword
%display desired
\maketitle

\section{Introduction}
Photon pair sources based on spontaneous parametric processes have
represented a crucial enabling technology for fundamental tests of
quantum mechanics~\cite{zeilinger99} and for the implementation of
quantum-information processing protocols~\cite{kok07}.    An
important distinction among spontaneous parametric photon-pair
sources is whether they are based on second-order non-linearities in
crystals~\cite{burnham70}, or on third-order non-linearities, often
in optical fibers~\cite{sharping01,Lee06}.  In the first case, we
refer to the process as spontaneous parametric downconversion (SPDC)
while in the second case we refer to the process as spontaneous
four-wave mixing (SFWM). The fact that two pump photons are
annihilated per photon pair generation event in the case of SFWM
rather than only one, in the case of SPDC, represents the key
underlying difference between these two processes.   This essential
difference can make SFWM sources superior in terms of a greater
ability to engineer the photon-pair properties~\cite{garay07}, and
in terms of a different dependence of the emitted flux on certain
experimental parameters which favors bright photon-pair sources.

Of course, an important consideration in the design and
implementation of photon-pair sources is the emission flux, or
equivalently, the conversion efficiency.    On the one hand, the
ability to compare the expected flux in a number of different
experimental situations is an important tool for the design of
specific sources and experiments.  On the other hand, our
understanding of a photon-pair source is not complete without a full
appreciation of the dependence of the conversion efficiency on all
relevant experimental parameters.   The motivation behind this paper
is to present specific analytic expressions for  the conversion
efficiency expected in the spontaneous four wave mixing process in
optical fibers.  While we restrict our attention to co-polarized
SFWM (i.e. we assume the same polarization for all four fields) and
likewise we restrict our treatment to cases where all four fields
propagate in the fundamental fiber mode, we consider a number of
other source variations.  Specifically, we consider both pulsed and
monochromatic pumps, as well as both degenerate and non-degenerate
pumps.   We derive expressions for the conversion efficiency in the
form of integrals, which where possible we take to closed analytic
form under certain approximations.   We compare values derived from
numerical integration of the conversion efficiency expressions
(without resorting to approximations) to corresponding values
derived from expressions in closed analytic form. We also note that
this work could be extended in a straightforward manner to also
incorporate the cases of cross-polarized SFWM, and of the SFWM
process in bi-refringent fibers.

Let us note that for second-order non-linear processes which are
stimulated in nature, such as second harmonic generation, the
emitted flux scales as the square of the incident pump power, and
scales linearly with the pump bandwidth~\cite{Arbore}.   This is a
result of the fact that two pump photons are combined to generate
each second-harmonic photon.   In the case of SFWM, even though the
process is spontaneous in nature, two pump photons are likewise
involved in every photon-pair generation event.  This leads to the
same dependence of emitted flux on pump power and pump bandwidth,
for spontaneous four wave mixing in a third-order non-linear medium,
as compared to (stimulated) second harmonic generation in a
second-order non-linear medium.  This is to be contrasted with SPDC
for which the emitted flux scales linearly with pump power and is
constant (within the phasematching bandwidth) with respect to the
pump bandwidth.  The fact that in some respects SFWM sources behave
essentially as a stimulated process would in a second-order
non-linear medium, coupled with long interaction lengths possible,
favors SFWM over SPDC sources in terms of the attainable photon-pair
flux. As a concrete illustration, in a remarkable recent SPDC
experiment~\cite{Krischek10}, despite extensive source optimization
the observed photon-pair flux is $\sim 500$ times lower compared to
a representative SFWM experiment~\cite{fulconis07}, when computing
the flux per unit pump power and per unit emission bandwidth.

While some previous works have analyzed the emitted flux in the SFWM
process~\cite{chen05, alibart06, brainis09}, in this paper we aim to
present a unified approach leading to explicit conversion efficiency
expressions, together with corresponding numerical simulations,
valid for the pulsed- and monochromatic-pumps regimes, as well as
for the degenerate- and non-degenerate-pumps configurations.

%The dependence on the non-linear medium length, however, is linear both for SPDC and for SFWM in contrast to second harmonic generation where it is quadratic.

\section{Derivation of the rate of emission}

In this paper we study the process of spontaneous four-wave mixing in optical fibers, in which nonlinear phenomena originate from the third-order susceptibility $\chi^{(3)}$. In this process, two photons (one from each of two pump fields $E_1$ and $E_2$) can be jointly annihilated to give rise to the emission of a photon-pair comprised of one photon in the signal mode $\hat{E}_s$ and one photon in the idler mode $\hat{E}_i$.   In our analysis we assume that all fields propagate in the same direction along the fiber (which defines the $z$-axis), and in the fundamental transverse mode supported by
the fiber. The
electric fields can be written  in the form $E_\mu=(E_{\mu}^{(+)}+E_{\mu}^{(-)})/2$;
the superscripts $^{(+)}$/$^{(-)}$ denote the  positive-/negative-frequency parts of the electric field. While we assume that all four participating fields are linearly polarized, parallel to the $x$-axis,
our analysis could be adapted to cross-polarized spontaneous four wave mixing processes.  In this paper,
we specifically focus on the rate of emission of photon-pair sources based on spontaneous four wave
mixing in optical fiber.

It can be shown that the SFWM process is governed by the following Hamiltonian

\begin{align}
\label{Hamilt}\hat{H}(t)&=\frac{3}{4}\epsilon_o\chi^{(3)}\nonumber\\& \times\!\int\!\! d^3 \textbf{r} E_1^{(+)}(\textbf{r},t)E_2^{(+)}(\textbf{r},t)\hat{E}_s^{(-)}(\textbf{r},t)\hat{E}_i^{(-)}(\textbf{r},t),
\end{align}

\noindent where the integration is carried out over the portion of the nonlinear medium which is illuminated by the pump fields, and $\epsilon_o$ is the vacuum electrical permittivity.

The quantized signal and idler fields can be written in the form

\begin{align}
\label{Ecuant}\hat{E}^{(+)}(\textbf{r},t)&=i\sqrt{\delta k}f(x,y)\nonumber\\&\times \sum_{k}\mbox{exp}\left[-i(\omega t-kz)\right]\ell(\omega) \hat{a}(k),
\end{align}

\noindent where the angular frequency $\omega$ is a function of $k$, as defined by the dispersion relation. $\delta k=2\pi/L_Q$ is the mode spacing, written in terms of the quantization length $L_Q$.  Function $\ell(\omega)$ is given as follows

\begin{equation}
\label{ldek}\ell(\omega)=\sqrt{\frac{\hbar \omega}{\pi\epsilon_o n^2(\omega)}},
\end{equation}

\noindent in terms of the (linear) refractive index of the nonlinear medium $n(\omega)$ and of Planck's constant $\hbar$. In Eq.~(\ref{Ecuant}), $\hat{a}(k)$ is the annihilation operator associated with the fundamental propagation mode in the fiber, and $f(x,y)$  represents the transverse spatial distribution of the field, which is normalized so that $\int\!\!\int|f(x,y)|^2\,dxdy=1$, and which is here approximated to be frequency-independent within the bandwidth of signal and idler modes.

In our analysis, we assume that the two pumps can be well-described by classical fields, expressed in terms of their Fourier components as

\begin{align}
\label{pump}E_{\nu}^{(+)}(\textbf{r},t)&=A_{\nu}f_{\nu}\left(x,y\right)\nonumber\\&\times\int\!\!d\omega\alpha_{\nu}(\omega)\,\mbox{exp}\left[-i\left(\omega t-k(\omega)z\right)\right],
\end{align}

\noindent where, for each of the two pump fields ($\nu=1,2$):  $A_{\nu}$ is the amplitude, $\alpha_{\nu}(\omega)$ is the spectral envelope with normalization $\int\!\!d\omega|\alpha_{\nu}(\omega)|^2=1$ and $f_{\nu}(x,y)$ is the transverse spatial distribution. Functions $f_{\nu}(x,y)$ are approximated to be frequency-independent within the spectral width of the pump pulses and exhibit the same normalization as their signal and idler counterparts (see Eq.~(\ref{Ecuant})).  It can be shown that $A_{\nu}$ is related to pump peak power, $P_{\nu}$, according to the relation

\begin{equation}
\label{ampl}A_{\nu}=\left[\frac{2P_{\nu}}{\epsilon_ocn_{\nu}\left|\int\! d\omega\alpha_{\nu}(\omega)\right|^2}\right]^{1/2},
\end{equation}

\noindent in terms of  $n_{\nu}\equiv n(\omega^o_{\nu})$, where $\omega^o_{\nu}$ represents the carrier frequency for pump $\nu$.

By replacing Eqs.~(\ref{Ecuant}) and (\ref{pump}) into Eq.~(\ref{Hamilt}), and following a standard perturbative approach \cite{mandel}, it can be shown that the two-photon state produced by spontaneous four-wave mixing is given by $|\Psi\rangle=|0\rangle_s|0\rangle_i+\zeta|\Psi_2\rangle$, where $|\Psi_2\rangle$ is the two-photon component of the state

\begin{equation}
\label{quantest2}|\Psi_2\rangle=\sum_{k_s}\!\sum_{k_i}G_k(k_s,k_i)\,\hat{a}^\dag(k_s)\hat{a}^\dag(k_i)
|0\rangle_s|0\rangle_i,
\end{equation}

\noindent written in terms of the joint amplitude $G_k(k_s,k_i)$ and a constant $\zeta$, related to the conversion efficiency

\begin{align}
\label{zeta} \zeta&=i\frac{3(2\pi)\chi^{(3)}\epsilon_oL\delta k}{4\hbar}\nonumber\\&\times A_1A_2\!\!\int\!\!dx\!\!\int\!\!dyf_1(x,y)f_2(x,y)f_s^*(x,y)f_i^*(x,y).
\end{align}

The function $G(\omega_s,\omega_i)= \ell(\omega_s)\ell(\omega_i)F(\omega_s,\omega_i)$ results from writing $G_k(k_s,k_i)$ in terms of frequencies rather than wavenumbers and represents the joint spectral amplitude, written in terms of the function $F(\omega_s,\omega_i)$ given by

\begin{align}
\label{JSA}F(\omega_s,\omega_i)=\int&\!d\omega\,\alpha_1(\omega)\alpha_2(\omega_s+\omega_i-\omega)\times\nonumber\\&\mbox{sinc}\!\left[\frac{L}{2}\Delta k(\omega,\omega_{s},\omega_{i})\right]e^{i\frac{L}{2}\Delta k(  \omega,\omega_{s},\omega_{i})}.
\end{align}

Note that the spectral dependence of $\ell(\omega)$ (see Eq.~(\ref{ldek})) tends to be slow over the frequency range of interest.  If this dependence is neglected~\cite{Rubin94},  the photon pair spectral properties are fully determined by function $F(\omega_s,\omega_i)$, which from this point onwards we refer to as the joint spectral amplitude function.   In Eq.~(\ref{JSA}), $\Delta k(\omega,\omega_{s},\omega_{i})$ represents the phasemismatch, given by

\begin{align}
\Delta k\left(  \omega,\omega_{s},\omega_{i}\right) &=k\left(  \omega
\right)  +k\left(  \omega_{s}+\omega_{i}-\omega\right) \nonumber\\& -k\left(
\omega_{s}\right)  -k\left(  \omega_{i}\right)  -\left(\gamma_1P_{1}+\gamma_2P_{2}\right)  , \label{eq: delk}%
\end{align}

\noindent which includes a nonlinear contribution ($\gamma_1P_{1}+\gamma_2P_{2}$) derived from self/cross-phase modulation \cite{garay07}, where $\gamma_{\nu}$ is the nonlinear coefficient given by

\begin{equation}
\gamma_{\nu}=\frac{3\chi^{(3)}\omega_{\nu}^o}{4\epsilon_oc^2n_{\nu}^2A^{\nu}_{eff}}\label{gamj}.
\end{equation}

In the above expression, we have defined $n_{\nu}\equiv n(\omega_{\nu}^o)$ and the effective area $A_{eff}^{\nu}\equiv [\int\!\int\!dxdy|f_{\nu}(x,y)|^4]^{-1}$, in terms of the carrier frequency $\omega_{\nu}^o$ for pump-mode $\nu$~\cite{agrawal2007,NLshift}.

For ease of calculating the emitted flux for fiber-based SWFM sources we assume that the photon pairs, once generated in the fiber of length $L$ considered in our calculation,  propagate along a continuation of this fiber to the detectors, in such a way that no further photon pairs are produced.   In a realistic experiment this could be achieved by suppressing pump photons through the use of appropriate spectral filters. Thus, the (linear) optical properties of the fiber through which the photon pairs propagate in order to reach the detectors is assumed to be identical to those of the fiber length where generation takes place.   Likewise, we assume that the photon pairs are split into separate spatial signal and idler modes (e.g. with a fiber-based beam splitter) and we refer to the count rate of an individual mode (e.g. the signal mode) as the source brightness. In order to proceed with our analysis, we are interested in the expectation value of the signal-mode energy density given by

\begin{equation}
\label{enerden}
u(\textbf{r},t)=\frac{1}{2}\langle\Psi_2|\hat{E}^{(-)}_s(\textbf{r},t)\hat{D}^{(+)}_s(\textbf{r},t)|\Psi_2\rangle,
\end{equation}

\noindent where $\hat{D}_s(\textbf{r},t)=\epsilon\hat{E}_s(\textbf{r},t)$ is the signal-mode electric displacement operator,  $\epsilon$ is the medium permittivity and $\hat{E}_s$ is given according to Eq.~(\ref{Ecuant}). Nevertheless, since propagation occurs along the z axis, it is convenient to calculate the linear energy density $u_z(z,t)$ by integrating $u(\textbf{r},t)$ over the transverse coordinates $x$ and $y$. Consequently, replacing Eqns.~(\ref{Ecuant}) and (\ref{quantest2}) into Eq.~(\ref{enerden}) we can show that

\begin{align}
\label{densyEnerz}u_z(z,t)=\,\vartheta\!\!\int&\!\!dk_s\!\!\int\!\!dk'_s\!\!\int\!\!dk_i\,\ell[\omega(k_s)]\ell_D[\omega(k'_s)]\nonumber\\&\times G^{\ast}(k_s,k_i)G(k'_s,k_i)\nonumber\\&\times e^{-i[\omega(k'_s)-\omega(k_s)]t}e^{-i(k_s-k'_s)z},
\end{align}

\noindent where $\vartheta=2|\zeta|^2/(\delta k)^2$ (note that $\zeta$ is linear in $\delta k$ so that $\vartheta$ is constant with respect to $\delta k$) and is explicitly given by

\begin{align}
\label{cte2}
\vartheta&=\frac{2^3(2\pi)^2\epsilon_o^2n_1n_2c^2}{\hbar^2\omega_1^o\omega_2^o}\frac{L^2\gamma^2P_1P_2}{|\!\int\!\!d\omega\alpha_1(\omega)|^2|\!\int\!\!d\omega\alpha_2(\omega)|^2},
\end{align}

\noindent and $\ell_D(\omega)=\epsilon_o n^2\ell(\omega)$. The
linear energy density, see (Eq.~(\ref{densyEnerz})), has been
written in terms of the nonlinear coefficient $\gamma$~\cite{gammas}
(which is different from $\gamma_1$ and $\gamma_2$  of
Eq.~(\ref{gamj})), defined as

\begin{equation}
\label{gamma}\gamma= \frac{3\chi^{(3)}\sqrt{\omega_1^o\omega_2^o}}{4\epsilon_oc^2n_1n_2A_{eff}},
\end{equation}

\noindent where $A_{eff}$ is the effective interaction area among the four fields given by

\begin{equation}
\label{aeffe}A_{eff}=\frac{1}{\int\!\!dx\!\!\int\!\!dyf_1(x,y)f_2(x,y)f_s^*(x,y)f_i^*(x,y)}.
\end{equation}

Note that the above expression for $A_{eff}$ takes into account the
normalization assumed for the transverse spatial distribution of
each of the four modes which participate in the process of SFWM.

%Note that in this paper we neglect the frequency dependence of $\gamma$; we use the value of $\gamma$ which results from evaluation at the carrier frequencies of the two pump modes.
%From this point onward, we use the approximation that the transverse spatial distribution [functions $f(x,y)$] is the same for these four modes \cite{agrawal2007}.

In  Eq.~(\ref{densyEnerz}) $k$-vector sums have been  converted to integrals, which can be done in
the limit $L_Q\rightarrow\infty$, so that $\delta k\sum_k\rightarrow\int\!\!dk$. This equation gives the signal-mode linear energy density.  The total signal-mode energy can be obtained by integrating $u_z(z,t_0)$, evaluated at a given time $t_0$, over coordinate $z$ within the spatial extent of the generated bi-photon wavepackets. For this calculation, we consider two cases: i) pulsed pump fields, and ii) monochromatic pump fields, which are addressed in the following two subsections.

\subsection{Pulsed pump configurations}

%In the cases of SFWM with pulsed pumps the signal-mode energy generated per pump pulse can be calculated as $U_{s}=\int^{\infty}_{-\infty}\!dz\,u_z(z,t)$, where $u_z(z,t)$ is the linear energy density (see Eq.~(\ref{densyEnerz})).

In order to carry out further calculations for the specific case of
pulsed pumps, we limit our treatment to pump fields with a Gaussian
spectral envelope, which can be written in the form

\begin{equation}
\label{envSpec}\alpha_{\nu}(\omega)=\frac{2^{1/4}}{\pi^{1/4}\sqrt\sigma_{\nu}}\,\exp
\left[-\frac{(\omega-\omega_{\nu}^o)^2}{\sigma_{\nu}^2}\right],
\end{equation}

\noindent where $\omega_{\nu}^o$ represents the central frequency and $\sigma_{\nu}$ defines the  bandwidth. Replacing Eqs.~(\ref{ldek}), (\ref{JSA}), (\ref{cte2}) and (\ref{envSpec}) into Eq.~(\ref{densyEnerz}) we obtain the following expression for the SFWM signal-mode energy produced by an isolated mode-$1$ pump pulse interacting with an isolated mode-$2$ pump pulse

\begin{align}
\label{Enerfwm2}U_{s} &=\int\limits_{-\infty}^\infty d z u_z(z,t_0)=\frac{2^6\hbar c^2n_1n_2}{\pi^2\omega_1^o\omega_2^o} \frac{L^2\gamma^2P_1P_2}{\sigma_1^2\sigma_2^2} \nonumber\\&\times \int\!\!d\omega_s\!\!\int\!\!d\omega_i\frac{\omega_s^2k^{(1)}(\omega_s)}{n^2(\omega_s)}\frac{\omega_i k^{(1)}(\omega_i)}
{n^2(\omega_i)}\left|f(\omega_s,\omega_i)\right|^2,
\end{align}

\noindent in terms of a version of the joint spectral amplitude (see
Eq.~(\ref{JSA})) defined as
$f(\omega_s,\omega_i)=(\pi\sigma_1\sigma_2/2)^{1/2}F(\omega_s,\omega_i)$,
which does not contain factors in front of the exponential and sinc
functions so that all factors appear explicitly in
Eq.~(\ref{Enerfwm2}).  Note that the signal-mode energy density has
appreciable values within the overlap region between the two pump
pulses along $z$; because we are considering for this calculation a
single isolated pulse for each of the two pump modes,  we have
extended the integration limits to $\pm \infty$ in
Eq.~(\ref{Enerfwm2}).  In the derivation of this equation, integrals
over $k_s$ and $k_i$ were transformed to frequency integrals through
the relation $dk=k^{(1)}(\omega)d\omega$, where $k^{(1)}(\omega)$
represents the first frequency derivative of $k(\omega)$.

%\begin{align}
%\label{JSA1}f(\omega_s,\omega_i)=\int&\!\!d\omega\,\exp\left[-\frac{(\omega-\omega_1^o)^2} {\sigma_1^2}\right]\exp\left[-\frac{(\omega_s+\omega_i-\omega-\omega_2^o)^2}{\sigma_2^2}\right] \times \nonumber\\&\mbox{sinc}\!\left[\frac{L}{2}\Delta k(  \omega,\omega_{s},\omega_{i})\right]\!\exp \left[ i\frac{L}{2}\Delta k(  \omega,\omega_{s},\omega_{i}) \right].
%\end{align}

In order to calculate the number of signal-mode photons generated
per mode-$1$/mode-$2$ pump pulse pair, we first express $U_{s}$ in
terms of the signal-mode spectral energy density, $u_s(\omega_s)$,
defined such that $U_{s}=\int\!\!d\omega_su_s(\omega_s)$. The
corresponding spectral photon number density is then given by
$\mathscr{N}_s(\omega_s)=u_s(\omega_s)/(\hbar\omega_s)$ and finally
the total emitted-photon number is obtained as
$N_s=\int\!\!d\omega_s\mathscr{N}_s(\omega_s)$.

We are interested in calculating the conversion efficiency in the
co-polarized SFWM process, which we define as $\eta=N_s/N_p$, where
$N_p=N_1+N_2$; here, $N_\nu$  is the number of photons per pump
pulse for each of two pump modes (with $\nu=1,2$). For sufficiently
narrowband pump pulses, it is acceptable to write
$N_\nu=U_\nu/(\hbar \omega_\nu^o)$, where $U_\nu$ is the energy per
pulse in mode $\nu$.  In this case, we arrive at the following
expression for $N_\nu$

\begin{equation}
\label{photpump}N_{\nu}=\frac{\sqrt{2\pi} P_{\nu}}{\hbar\omega_{\nu}^o\sigma_{\nu}}.
\end{equation}

The photon-pair conversion efficiency can then be written as

\begin{align}
\label{etapul}\eta=&\frac{2^8\hbar^2 c^2n_1n_2}{(2\pi)^{3}} \frac{L^2\gamma^2N_1N_2}{\sigma_1\sigma_2(N_1+N_2)}  \nonumber\\&\times \int\!\!d\omega_s\!\!\int\!\!d\omega_i\frac{\omega_sk_s^{(1)}}{n_s^2}\frac{\omega_ik_i^{(1)}}{n_i^2}\left|f(\omega_s,\omega_i)\right|^2.
\end{align}

Through Eq.~(\ref{etapul}), we may gain an understanding of the
dependence of the conversion efficiency on various experimental
parameters, including  the fiber length $L$, the pump peak powers
($P_1$ and $P_2$) and the pump bandwidths ($\sigma_1$ and
$\sigma_2$). Besides these parameters, the conversion efficiency of
course also exhibits a dependence on fiber dispersion properties
through the phasemismatch (see Eq.~(\ref{eq: delk})).   From
Eq.~(\ref{etapul}), it is clear that $\eta$ varies quadratically
with the nonlinear coefficient $\gamma$, which implies that it has
an inverse fourth power dependence on the transverse mode radius
\cite{agrawal2007}. This means that, in general, a small core radius
leads to large rates of emission (note that this trend may be
reverted for sufficiently narrow fibers for which the mode radius
can increase as the core radius is reduced \cite{tong04, foster04}).

The dependence on pump peak power is clear from Eq.~(\ref{etapul}),
where $N_\nu$ is linear in $P_\nu$ according to
Eq.~(\ref{photpump}).  As expected, $N_s$ exhibits a quadratic
dependence on the pump power (or, alternatively, $\eta$ exhibits a
linear dependence on the pump power), which is more evident for
degenerate pumps, for which $P_1=P_2\equiv P$.   This behavior
represents an important difference with respect to photon-pair
sources based on SPDC in second-order nonlinear crystals, for which
$N_s$ is linear in pump power.    In fact, a quadratic dependence of
the generated power on the pump power, in the case of second-order
nonlinear optics, is associated with stimulated processes such as
second harmonic generation rather than with spontaneous processes.
This quadratic pump-power scaling represents a clear advantage of
the process of SFWM over SPDC, in terms of the attainable
photon-pair flux, for the design of bright photon-pair sources. Note
that because the phasemismatch has an additive term which is linear
in $P_1$ and $P_2$, for large enough pump powers there can be a
deviation from this stated quadratic dependence.

In common with SPDC, a concern with SFWM is that for sufficiently
high conversion efficiencies, multiple photon pairs can be generated
at a given time. This represents a limitation, since many
experiments rely on the emission of individual photon pairs, i.e.
which can be well isolated from other photons both spatially and
temporally.    Thus, for example, the existence of multiple-pair
amplitudes implies that a detection event (with a non-photon-number
resolving detector), say in the idler mode of an SPDC source, cannot
herald a true single photon in the signal mode of the source. In
practice, this means that the conversion efficiency (which depends
on experimental parameters such as the nonlinearity and the pump
power) must be limited so that the probability of multiple pair
emission remains negligible. Let us note that other physical systems
(e.g. biexcitonic decay in quantum dots~\cite{benson00}) are capable
of true photon pair emission. However, SPDC and SFWM remain highly
flexible platforms for photon pair emission, with entanglement
characteristics which can be tailored according to the requirements
of specific applications.

% and atomic ensembles~\cite{}

The dependence of the conversion efficiency on the fiber length $L$
and on the pump bandwidths ($\sigma_1$ and $\sigma_2$) is not as simple to deduce from Eq.~(\ref{etapul}),
compared to the pump power dependence,
because these parameters are implicit in the joint spectral function (see Eq.~(\ref{JSA})) which is given by a convolution-type integral.   In general, as $L$ increases the
joint spectral intensity $|f(\omega_s,\omega_i)|^2$ tends to exhibit a width in the space of generated
frequencies which scales as $L^{-1}$.  Equation~(\ref{etapul}) then tells us that the conversion efficiency
tends to be linear in $L$.  However, as we will see below, certain situations (such as non-degenerate
pumps) can lead to a deviation from this linear behavior.
Of course, a natural advantage of spontaneous four-wave mixing over spontaneous parametric
downconversion sources is that the interaction length can be increased easily, simply by increasing
the fiber length  (while, usually, non linear crystals tend to be limited in length to a scale of mm or cm).

While in general it is not possible to find a closed analytic expression for $\eta$, we will show in the next section
that this becomes possible under certain approximations.

\subsection{Pulsed pumps: closed analytic expressions} \label{anaexp}

In Ref.~\cite{garay07}, we showed that it is possible to derive a closed analytic expression for the
joint spectral amplitude (see Eq.~(\ref{JSA})), if we resort to a linear
approximation of the phasemismatch.   Specifically, this approximation involves writing the
phasemismatch (see Eq.~(\ref{eq: delk})) as a first-order Talyor expansion in the frequency detunings $\omega_\mu-
\omega_\mu^o$ (for $\mu=s,i$) where $\omega_\mu^o$ represents the signal and idler frequencies for which perfect
phasematching is obtained.  Here, we exploit this approximation in order to obtain analytic expressions
for the conversion efficiency.

We start by defining the function  $h(\omega_s,\omega_i)$, which constitutes a factor in the
integrand of the expression for the conversion efficiency (see Eq.~(\ref{etapul})).

\begin{equation}
\label{hwswi}h(\omega_s,\omega_i)=\frac{ \omega_s\omega_ik_s^{(1)}(\omega_s)k_i^{(1)}(\omega_i)}
{ n_s^2(\omega_s)n_i^2(\omega_i)}.
\end{equation}

Next, we assume that $h(\omega_s,\omega_i)$ varies slowly over the spectral range of interest,
so that we can consider it to be a constant when evaluating the integral in Eq.~(\ref{etapul}).
We analyze first the general case in which the pumps are non-degenerate, i.e. where they can differ
both in central frequency ($\omega_1^o$ and $\omega_2^o$) and in bandwidth
($\sigma_1$ and $\sigma_2$). Following the treatment presented in section 2 of Ref.~\cite{garay07}
it is possible to find an analytic
expression for the conversion efficiency $\eta$ ($NDP$ below refers to non-degenerate pumps)

\begin{align}
\label{etaNDP}\eta^{NDP}=&\frac{2^5\hbar^2 c^2n_1n_2\gamma^2N_1N_2}{(N_1+N_2)}
\nonumber\\&\times\frac{\mbox{erf}\left[(\sqrt{2}B)^{-1}\right]h(\omega_s^o,\omega_i^o)}
{\big|k_1^{(1)}-k_2^{(1)}\big|\big|k_i^{(1)}-k_s^{(1)}\big|},
\end{align}

\noindent where $k^{(1)}_{\mu}=k^{(1)}(\omega_{\mu}^o)$ (with $\mu=1,2,s,i$), where
$\mbox{erf}(x)$ is the error function and where we have defined the parameter $B$ as\cite{garay07}

\begin{equation}
\label{B}B=\frac{\left(  \sigma_{1}^{2}+\sigma_{2}^{2}\right)  ^{1/2}}{\sigma_{1}\sigma_{2}
L[k_1^{(1)}-k_2^{(1)}] }.
\end{equation}

Note that the only
dependence of the conversion efficiency on the pump bandwidth and the fiber length
is through the $B$ parameter in the argument of the error function. The error function in Eq.~(\ref{etaNDP})  implies that
$\eta^{NDP}$ exhibits a saturation behavior when $L$ or $\sigma_{1,2}$ vary, governed
by the group-velocity mismatch between the two pumps. For example, we expect that for fixed pump bandwidths
the
conversion efficiency reaches a plateau at the point where the two pump pulses no longer overlap in time
due to their different group velocities.

Let us now consider the degenerate pumps limit of Eq.~(\ref{etaNDP}).  In this case, making
$\sigma_1=\sigma_2 \equiv \sigma$ and $P_1=P_2 \equiv P$,
in the limit $\omega_1^o\rightarrow\omega_2^o \equiv \omega^o$,  Eq.~(\ref{etaNDP}) reduces to

\begin{equation}
\label{etaDP}\eta^{DP}=\frac{2^4\hbar^2 c^2n^2(\omega^o) L\sigma N\gamma^2}{\sqrt{\pi}\big|k_i^{(1)}-k_s^{(1)}\big|}h(\omega_s^o,\omega_i^o),
\end{equation}

\noindent where $DP$ denotes degenerate pumps. In the above equation it is possible to observe that the SFWM conversion efficiency increases
linearly with fiber length and pump pulse bandwidth, at least within the
frequency range in which the linear approximation for the phase mismatch is valid.

Note that in both the degenerate pumps and non-degenerate pumps cases, the emitted flux is inversely
proportional to the group-velocity mismatch between signal and idler modes.   As $k_s^{(1)}$ approaches
$k_i^{(1)}$, the orientation of the phasematching function
in the generated frequencies space
$\{\omega_s,\omega_i\}$ approaches that of the pump envelope function~\cite{garay07}, with the implication that the emitted bandwidth increases, and consequently the
generated flux also
increases.   In the case where $k_s^{(1)}=k_i^{(1)}$ (which would result from making the signal
and idler frequencies degenerate), the linear approximation is no longer
sufficient;  second- and higher-order terms (not present within this approximation) prevent
the resulting divergence in Eq.~(\ref{etaNDP}).  From this analysis, it becomes clear that
sources with a small signal-idler spectral separation tend to exhibit
a considerably higher brightness than sources with a large signal-idler spectral separation.

%Even though the linear approximation used here is strictly valid in a small area
%of $\{\omega_s,\omega_i\}$ space, centered at frequencies in which perfect phase
%matching occurs, they represent an important tool to estimate the photon number
%generated by pulsed SFWM, with good accuracy due to $\Delta k$ is dominated by
%first-order terms in the Taylor series expansion. This affirmation is justified
%by the agreement between the analytical and numerical results that will be exposed later.

\subsection{Narrowband pump configurations}

In this section, we focus our attention on SFWM photon pair
sources involving pumps in the monochromatic limit, i.e. for which $\sigma_{1,2} \rightarrow 0$.
If the SFWM process takes place in a single transverse mode environment, such as
a single-mode fiber, factorability
is enforced on the transverse momentum degree of freedom, leaving frequency as the only
continuous-variable degree of freedom where entanglement may reside.  Let us note that SFWM sources based on
narrow-band pumps permit the emission of photon pairs which are
highly entangled in frequency~\cite{garay08}.  Here we present an analysis of the conversion efficiency
for this type of source.

%Now, we present a theoretical treatment that permits us to quantify the photon-pair number generated
%by SFWM with cw-pumps, and thus determine the physical characteristics that lead to high
%emission rates under this configuration. For a same pump average power, it is expected
%that the emitted flux in the limit of monochromatic pumps is less than that obtained
%by using a pulsed-pumps configuration. This is related to the fact that the energy
%generated by pulsed-SFWM is dependent on the peak power, which linearly increases
%with the pump bandwidth.

In order to proceed with the calculation we take the limit $ \sigma_1=\sigma_2 \equiv \sigma \rightarrow 0$ of the
linear energy density $u_z(z,t)$ (see Eq.~(\ref{densyEnerz}))~\cite{garay08}, obtaining the following time-independent
expression

\begin{align}
\label{Uz} u_z(z)=\vartheta_{cw}\!\!&\int\!\!dk_s\!\!\int\!\!dk_i\,\ell^3(\omega_s)
\ell_D(\omega_s)\ell^2(\omega_i)k_s^{(1)}\nonumber\\&\times|F_{cw}(\omega_s,\omega_i)|^2,
\end{align}

\noindent in terms of the joint amplitude function $F_{cw}(\omega_s,\omega_i)$ and of the parameter $\vartheta_{cw}$,
given by

\begin{align}
\label{JSA1cw} F_{cw}(\omega_s,\omega_i)&=\delta(\omega_s+\omega_i-\omega_{1}-\omega_{2})
\nonumber\\&\times\mbox{sinc}\left[\frac{L}{2}\Delta k_{cw}(\omega_s,\omega_i) \right]
\nonumber\\&\times e^{i\frac{L}{2}\Delta k_{cw}(\omega_s,\omega_i) }, \nonumber \\
\vartheta_{cw}&=\frac{2^3(2\pi)^2\epsilon_o^2n_1n_2c^2}{\hbar^2\omega_1\omega_2}L^2\gamma^2p_1p_2,
\end{align}

\noindent where $\Delta k_{cw}(\omega_s,\omega_i)$ is the phasemismatch, which is now a
function only of $\omega_s$ and $\omega_i$

\begin{align}
\label{Dkcw} \Delta
k_{cw}(\omega_s,\omega_i)&=k\left[\left(\omega_s+\omega_i+\omega_{1}
-\omega_{2}\right)/2\right]\nonumber\\&+k\left[\left(\omega_s+\omega_i-\omega_{1}+
\omega_{2}\right)/2\right]\nonumber\\& -k(\omega_s)-k(\omega_i)-(
\gamma_1p_1+\gamma_2p_2).
\end{align}

In Eq.~(\ref{JSA1cw}), $\omega_\nu$ represents the frequency and $p_\nu$ represents
the average power for each of the two pump modes (with $\nu=1,2$). Next, we define the
spectral linear energy density $\mathscr{U}_z(k_s,z)$ corresponding to an optical mode
with propagation constant $k_s$, such that $u_z(z)=\int\!\!dk_s \mathscr{U}_z(k_s,z)$. The
signal-mode energy reaching a detector placed at position $z=z_D$ during a time interval
$\Delta t$ is then given by $\mathscr{U}(k_s)=\int_{z'_D}^{z_D}dz\,\mathscr{U}_z(k_s,z)$,
where $z'_D=z_D-\Delta t/k_s^{(1)}$. Then,
the signal-mode energy reaching the detector from all modes $k_s$ is given by

\begin{equation}
\label{Ucw}U_{s}^{cw}=\int\!\!dk_s\,\mathscr{U}(k_s).
\end{equation}

From Eq.~(\ref{Ucw}),  following a treatment similar to that used for the pulsed-pumps case, it can be shown
that the number of photons reaching the detector during time $\Delta t$ is given by

\begin{align}
\label{numofFotcw} N_{cw}&=\frac{2^5n_1n_2c^2L^2\gamma^2p_1p_2 \Delta t}
{\pi\omega_1\omega_2}\nonumber\\&\times\int\!\!d\omega h(\omega,\omega_1+\omega_2-\omega)\mbox{sinc}^2[L\Delta k'_{cw}(\omega)/2],
\end{align}

\noindent where $\Delta k'_{cw}(\omega)=\Delta k_{cw}(\omega,\omega_1+\omega_2-\omega)$
and the function $h(\omega,\omega_1+\omega_2-\omega)$ is given according to Eq.~(\ref{hwswi}).

The energy launched into the fiber corresponding to pump mode $\nu$ (with $\nu=1,2$) during the time
interval $\Delta t$ is $U_\nu=p_\nu \Delta t$, so that the total photon number from two pumps in the time interval $\Delta t$ can be obtained as

\begin{equation}
\label{photpumpcw} N_{p,cw}=\frac{p_1\omega_2+p_2\omega_1}{\omega_1\omega_2} \frac{\Delta t}{\hbar},
\end{equation}

\noindent and therefore, the conversion efficiency $\eta_{cw}=N_{s}/N_{p,cw}$ in the process of SFWM
with monochromatic pumps is given by

\begin{align}
\label{eficw} \eta_{cw}&=\frac{2^5\hbar c^2 n_1n_2}{\pi}\frac{L^2\gamma^2p_1p_2}{p_1\omega_2+p_2\omega_1}\nonumber\\&\times\int\!\!d\omega h(\omega,\omega_1+\omega_2-\omega)\mbox{sinc}^2[L\Delta k'_{cw}(\omega)/2].
\end{align}

%Finally, we have an expression that permits us to calculate the photon-pair number per unit
% time generated by SFWM in the limit of monochromatic pumps. Note that as in the pulsed-pumps
%  case, $N_{cw}$ exhibits a quadratic behavior with pump power. The dependence with fiber length
%   is not explicit in Eq.~(\ref{numofFotcw}), however results obtained by numerical evaluation
%    of this equation show that the emitted photon-pair number linearly increases when $L$ is
%     increased for both degenerate and non-degenerate pump regimes, as it will be seen in a later section.

\section{Conversion efficiency in specific situations}\label{results}

In this section, we present the results of simulations of the
expected conversion efficiency as a function of various experimental
parameters: fiber length, pump power and pump bandwidth. We also
consider the dependence of the conversion efficiency on the type of
spectral correlations between the signal and idler photons.  Note
that for these simulations we have used the full two-photon state,
i.e. we have not resorted to approximations.   We compare these
simulations with plots derived from the analytic expressions for the
conversion efficiency presented in Sec.~\ref{anaexp}.    We include
in our analysis both, the degenerate and non-degenerate pump
configurations, as well as both, the pulsed and monochromatic pump
field regimes.  We assume that the SFWM process takes place in
photonic crystal fibers (PCFs), which have been widely used for the
experimental implementation of photon-pair sources \cite{rarity05,
fan05, cohen09}.   PCF's are typically characterized by higher
values of the nonlinearity coefficient $\gamma$,  due to a large
core-cladding dielectric contrast, as compared to typical
telecommunications fibers.  This leads to larger rates of emission
as is clear from Eqs.~(\ref{etapul}) and (\ref{eficw}). Furthermore,
PCF's permit the engineering of the fiber dispersion properties and
therefore of the resulting photon-pair properties \cite{garay07}.
In the following three subsections, we specifically consider a PCF
with a core radius of $r=0.97\mu$m and an air-filling fraction of
$f=0.91$; these values were chosen so that a zero dispersion point
exists at $\lambda=0.715 \mu$m. The fiber dispersion properties were
calculated through the step-index model proposed in
Ref.~\cite{Wong2005}.  We assume a repetition rate for the pulsed
pump modes of $f_r=80$MHz,  a fiber length of $L=0.5$m (except in
Sec.~\ref{Long} where we study the fiber length dependence), an
average pump power of $p=300\mu$W (except in Sec.~\ref{pot}, where
we study the pump power dependence), and a pump bandwidth of
$\sigma=3.0$THz (except in Sec.~\ref{bandw}, where we study the pump
bandwidth dependence).

For the degenerate-pumps configuration, we assume that the pump
pulses are centered at $\lambda_p=0.708\mu$m. This leads to a
numerically-calculated nonlinear coefficient, through
Eq.~(\ref{gamma}), of $\gamma=137$km$^{-1}$W$^{-1}$.  The pump peak
power, with a value of $4.5$W, derived from $\sigma=3$THz and
$p=300\mu$W, leads to generated signal and idler wave-packets
centered at $0.5759\mu$m and $0.9185\mu$m, respectively. For this
source configuration, the signal and idler frequencies are nearly
frequency-anticorrelated, with an orientation of the joint spectrum
in $\{\omega_s,\omega_i\}$ space of $\theta_{si}=-40^o$ with respect
to the $\omega_s$-axis\cite{garay07}.

For the non-degenerate pumps configuration, we assume that the two
pump fields can be obtained as the fundamental and second-harmonic
signal of the same laser system, thus facilitating the experimental
implementation. As discussed in Ref.~\cite{garay07} this
configuration permits the generation of signal and idler modes,
which are sufficiently distant (in frequency) from the pumps, so
that the signal and idler modes remain uncontaminated by photons
produced by spontaneous Raman scattering.  Specifically, we assume
that the pump central wavelengths are $\lambda_{1}^0=0.521\mu$m and
$\lambda_{2}^0=1.042\mu$m, which leads to a numerically-calculated
nonlinear coefficient for the SFWM interaction of
$\gamma=131$km$^{-1}$W$^{-1}$;  we note that this could be achieved
with a Yb:KGW laser~\cite{major06}.  The values of the source
parameters which we have assumed lead to signal and idler modes
centered at $\lambda_s^o=0.5826\mu$m and $\lambda_s^o=0.8600\mu$m,
respectively. Similarly to the degenerate pumps case above,
parameters were chosen so that this photon-pair source exhibits near
spectral anti-correlation, in this case with a joint spectrum
oriented at  $-41^o$ in $\{\omega_s,\omega_i\}$ space. It is worth
mentioning that variations in the pump peak power (which can result
from variations of the average power or bandwidth) can produce a
shift of the signal and idler central generation frequencies due to
the nonlinear term in Eq.~(\ref{eq: delk}) and could also produce
variations of the types of signal-idler spectral correlations.
However, we found that for the range of peak powers considered in
the following simulations, these changes are negligible.

\subsection{Fiber length dependence}\label{Long}

\begin{figure}[h!]
\begin{center}
\centering\includegraphics[width=6.5cm]{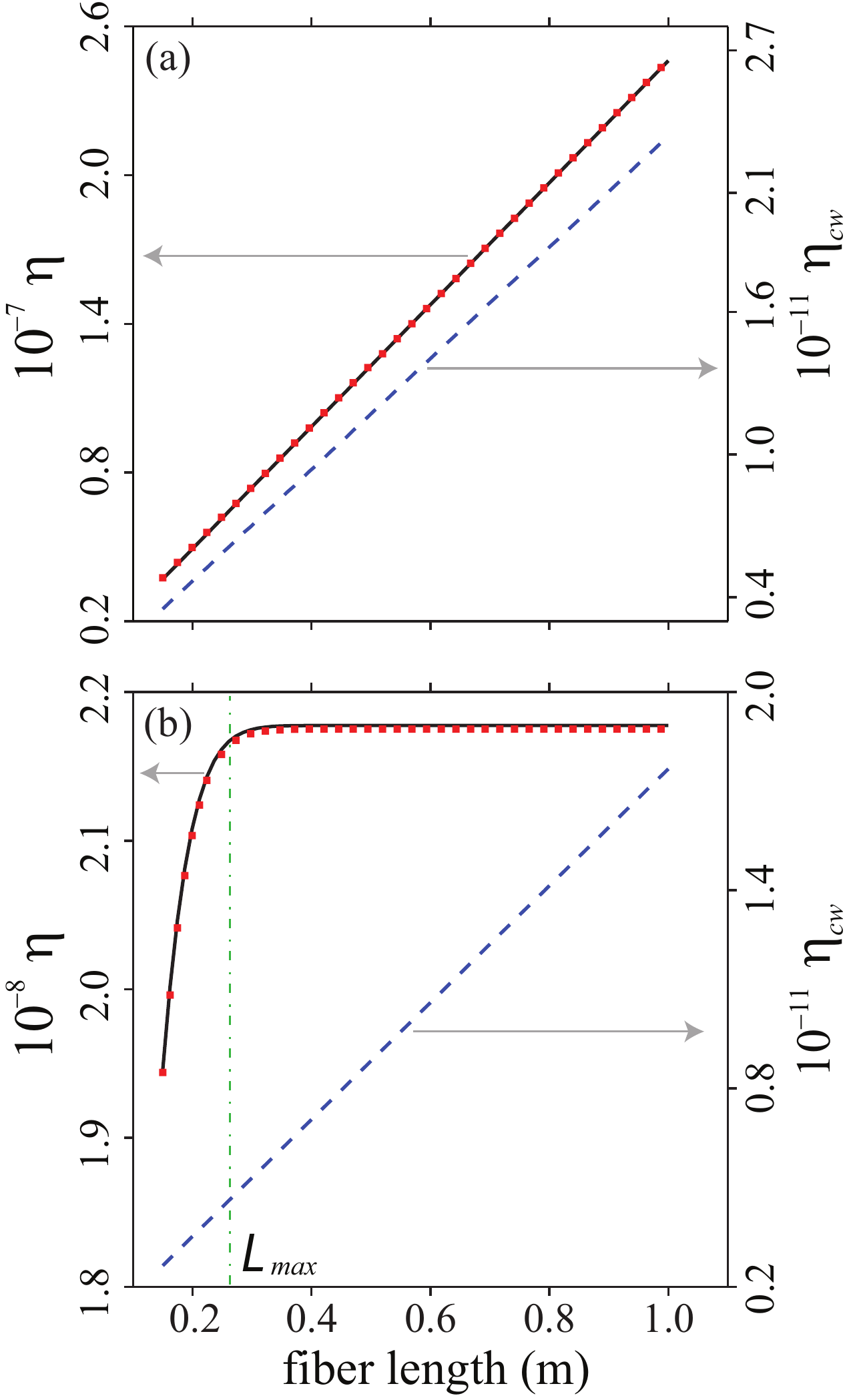}
\end{center}
\par
% Give a unique label
\caption{(Color online) Conversion efficiency  as a function of the fiber length for the pulsed and monochromatic pump regimes (denoted by $\eta$ and $\eta_{cw}$, respectively), and for the following configurations: (a) degenerate pumps, and (b) non-degenerate pumps. The red squares represent results obtained by numerical evaluation of Eq.~(\ref{etapul}). The solid-black line corresponds to results obtained from analytical expressions described in section \ref{anaexp}.} \label{varlong}
\end{figure}

Let us first consider the conversion efficiency for the photon-pair
sources described above as a function of the fiber length. For the
degenerate pumps configuration, the pump bandwidth which we have
assumed ($\sigma =3.0$THz) corresponds to a width in wavelength of
$\Delta\lambda=0.94$nm. The fiber length is varied between $0.15$
and $1.0$m. The results obtained by numerical evaluation of
Eq.~(\ref{etapul}) and those obtained from analytic expressions
(Eq.~(\ref{etaDP})) are presented in Fig.~\ref{varlong}(a). From
this figure, it can be seen that numerical and analytical results
are in good agreement over the full range of study, evidencing that
the analytical expression derived in the linear approximation of the
phase-mismatch is in fact an excellent approximation. As predicted
by Eq.~(\ref{etaDP}), the conversion efficiency exhibits a linear
dependence on fiber length. For the longest fiber considered
($L=1.0$m) approximately $5.3\times10^8$ photon pairs per second are
emitted.

Likewise, considering the same fiber length range and average pump
power as above, we evaluate the conversion efficiency $\eta_{cw}$ as
a function of the fiber length, in the monochromatic pump limit.
Results obtained by numerical evaluation of Eq.~(\ref{eficw}) are
represented in Fig.~\ref{varlong}(a) by a dashed-blue line. It can
be appreciated that the dependence of $\eta_{cw}$ on $L$  is, once
again, linear. For the source parameters which we have assumed and
for the longest fiber length considered, approximately
$5.0\times10^4$ photons/s are generated, which is much lower than
the emission rate attainable in the pulsed pumps regime for the same
fiber length.  This reflects a general trend: the conversion
efficiency is higher for pulsed-pumps than for monochromatic pumps,
because for SFWM (unlike for SPDC), the emission rate depends on the
peak power (which increases with increasing pump bandwidths) rather
than on the average pump power.
%, for which if the repetition rate is $f_r=80$ MHz, around $5.3\times10^8$ photons/s are generated.

For the non-degenerate pumps configuration, we assume that the two
pumps have the same bandwidth measured in frequency
$\sigma_1=\sigma_2=3$THz, corresponding to $\Delta\lambda_1=0.51$nm
and $\Delta\lambda_2=2.03$nm. It is also assumed that the pumps have
the same average power, so that  $P_1=P_2$. Fig.~\ref{varlong}(b)
shows the conversion efficiency as a function of fiber length, in
the range $0.15-1.0$m.  Clearly, there is an excellent agreement
between numerical and analytical results calculated from
Eqs.~(\ref{etapul}) and (\ref{etaNDP}), respectively. In this
figure, we can appreciate that if the fiber length exceeds a certain
value, denoted by $L_{max}$, the conversion efficiency reaches a
plateau. This effect is related to the pulsed nature of the
non-degenerate pump fields; the two fields experience different
group velocities in the fiber.  For a sufficiently long fiber, the
pump pulses no longer overlap temporally, and therefore photon-pair
generation ceases.  $L_{max}$ is defined as the value of $L$ which
makes the argument of the erf function in Eq.~(\ref{etaNDP}) equal
to $2$ (for which the erf function attains $99.5\%$ of its maximum
value), which leads to the following expression

\begin{equation}
\label{leff} L_{max}=\frac{2\sqrt{2}\sqrt{\sigma_1^2+\sigma_2^2}}{\sigma_1\sigma_2\left|k^{(1)}(\omega_1^o)-k^{(1)}(\omega_2^o)\right|}.
\end{equation}

This equation implies that for fixed pump bandwidths, the maximum
interaction length between the pumps is determined by the difference
of their reciprocal group velocities. Thus, pump fields with a
considerable spectral separation will tend to exhibit a short
maximum interaction length, which will be reflected in a low
emission rate in comparison to that attainable in a degenerate pumps
configuration, as is the case for the sources assumed for
Fig.~\ref{varlong}(a). In the particular case of
Fig.~\ref{varlong}(b), the maximum interaction length is
$L_{max}=0.263$m and $5.12\times10^7$ photon pairs per second are
generated for a fiber of length $L=L_{max}$.  Of course, if the
temporal duration of the pump pulses is increased, this will tend to
increase the maximum interaction length, since the mode-$1$ and
mode-$2$ pulses will remain temporally overlapped over a longer
length of fiber.  Results obtained for the monochromatic pumps
regime are also shown in Fig.~\ref{varlong}(b). In this case we
assumed the same pump power and fiber lengths as above.  It can be
appreciated that the conversion efficiency increases linearly with
fiber length over the full range of fiber length values. Indeed, for
non-degenerate monochromatic pumps there is no maximum interaction
length.

From these results, it is evident that pulsed pump regimes (both in
the degenerate and non-degenerate pumps configurations) lead to much
higher conversion efficiencies than monochromatic pump regimes. This
will also be clear from the discussion in Sec.~\ref{bandw}, where we
analyze the conversion efficiency as a function of pump bandwidth.
While non-degenerate pumps lead to a lower conversion efficiency
compared to degenerate pumps (for given pump bandwidths), in this
case the effect is much less drastic than for pulsed vs
monochromatic pumps.  Non-degenerate-pumps schemes offer some
advantages over degenerate-pumps schemes, despite the resulting
lower emission rates, especially in relation to the ability to
generate signal and idler photons away from the Raman gain bandwidth
of fused silica.

\subsection{Pump power dependence}\label{pot}

\begin{figure}[h!]
\begin{center}
\centering\includegraphics[width=6.5cm]{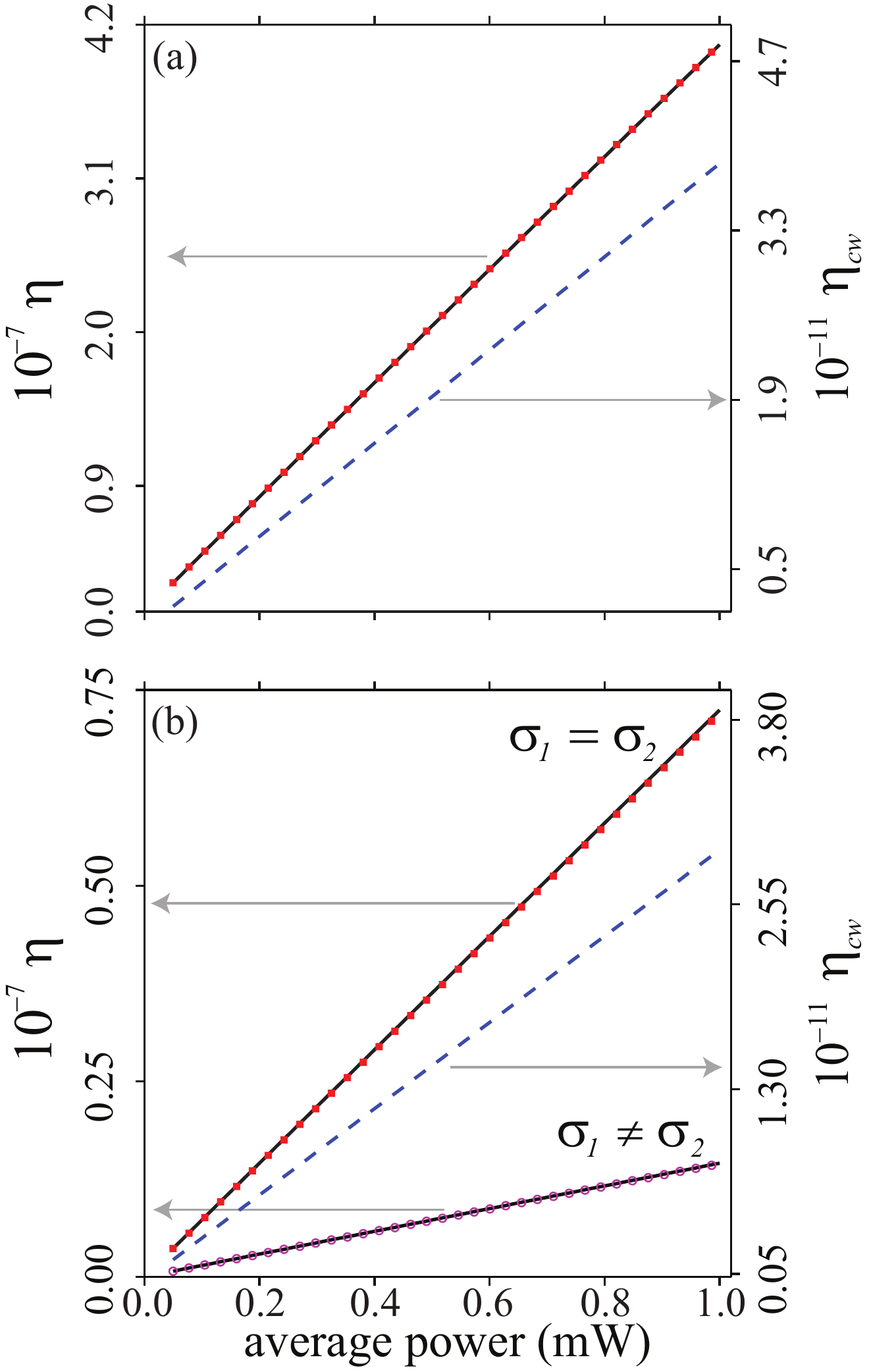}
\end{center}
\par
% Give a unique label
\caption{(Color online) Conversion efficiency as a function of average pump power, for pulsed and monochromatic pump regimes (denoted by $\eta$ and $\eta_{cw}$, respectively), and for the following configurations: (a) degenerate pumps and (b) non-degenerate pumps. The red squares and magenta circles are results obtained by numerical evaluation of Eq.~(\ref{etapul}), while the solid-black line corresponds to results obtained from analytical expressions described in section~\ref{anaexp}.} \label{varpot}
\end{figure}

We now turn our attention to the pump power dependence, while
maintaining the pump bandwidths fixed, of the conversion efficiency.
In order to compare sources with pulsed and monochromatic pumps, we
compute the conversion efficiency as a function of the
\emph{average} pump power together with, for the pulsed pumps case,
a certain value assumed for the repetition rate which characterizes
the pump-pulse train.   We consider the sources described above and
vary the average pump power between $0.05$ and $1.0$mW. Under these
conditions, for a repetition rate of $f_r=80$MHz, the pump peak
power varies within the range $0.75-15$W, without an appreciable
resulting variation of the spectral properties of the two-photon
state. Plots derived from our expressions (Eqs.~(\ref{etapul}) and
(\ref{etaDP})) are presented in Fig.~\ref{varpot}(a), from which it
is clear that the conversion efficiency $\eta$ is linear with pump
power.  Note that this linear dependence becomes quadratic, if the
emitted flux $N_s$, rather than the conversion efficiency, were to
be plotted vs average pump power.  This is a different behavior from
that observed for the generation of photons pairs through
spontaneous parametric downconversion (SPDC) in second-order
nonlinear materials, for which the conversion efficiency vs average
power is constant, while the emitted flux vs pump power is linear.
This is related to the fact that two pumps, rather than one, are
required for SFWM.  In fact, this represents one of the essential
advantages of SFWM over SPDC photon-pair sources:  while both
processes are spontaneous, in some respects such as pump power
dependence, SFWM behaves as a stimulated process exhibiting the same
pump power dependence as second-order non-linear processes such as
second harmonic generation.   At the highest average pump power
considered, Eq.~(\ref{etapul}) predicts the generation of
$2.89\times10^{9}$ photons pairs per second. For the corresponding
monochromatic pump case (see dashed-blue line in
Fig.~\ref{varpot}(a), the number of photon pairs generated for the
highest pump power considered becomes $2.75\times10^5$ per second.

For the case of pulsed, non-degenerate pumps we can analyze two
distinct configurations. On the one hand, the pumps can be
non-degenerate in frequency, with the same bandwidth so that
$\sigma_1=\sigma_2 \equiv \sigma$, which implies that the
corresponding peak powers are equal, i.e. $P_1=P_2$, if we assume
the same repetition rate for both pump modes. The numerical and
analytical results obtained from Eqs.~(\ref{etapul}) and
(\ref{etaNDP}) are shown in Fig.~\ref{varpot}(b) (red squares and
solid-black line, respectively). As shown, at the highest average
power considered, around $5.7\times10^8$ photons per second are
emitted. On the other hand, the bandwidth of the two pump modes can
be different. In this case, for the same average power and
repetition rate, the peak powers are no longer equal, i.e. $P_1 \neq
P_2$. In particular, we consider the limiting case where the pump
bandwidths are highly unequal, i.e.  $\sigma_1\ll\sigma_2$ (or
$\sigma_2\ll\sigma_1$).   In Fig.~\ref{varpot}(b) we present results
for $\sigma_1=0.1$THz and $\sigma_2=3.0$THz, where we assume the
same fiber length as above (numerical: magenta circles and
analytical: solid-black line). It can be appreciated that these
unbalanced pump bandwidths lead to an important reduction in the
conversion efficiency $\eta$ (while maintaining the average pump
power constant). The highest average pump power considered results
in $1.1\times10^8$ photon pairs emitted per second. This behavior
can be understood from Eqs.~(\ref{etaNDP}) and (\ref{B}), from which
we can show that the condition $\sigma_1\ll\sigma_2$ implies that
$\mbox{erf}[1/\sqrt{2}B]\ll1$. As in previous cases, we can observe
in Fig.~\ref{varpot} that analytical results are in excellent
agreement with numerical ones.

%Note that under these conditions, the type of photon-pair spectral correlations remains essentially unchanged as the pump power is varied. In this specific case, the fiber length exceeds the maximum interaction length $L_{max}=0.263$m.

For comparison, the conversion efficiency derived from
Eq.~(\ref{eficw}) in the monochromatic pumps regime is also shown in
Fig.~\ref{varpot}(b) (dashed-blue line). It can be appreciated that
the conversion efficiency in this configuration is several orders of
magnitude lower than for the pulsed pumps case.  Assuming $p_1=p_2$,
the emission rate at the highest pump power considered is
$2.3\times10^5$ photon pairs per second. This is to be compared with
a corresponding emission rate of $5.7\times10^8$ photon pairs per
second in the (non-degenerate) pulsed pumps regime, assuming the
same average pump powers and a repetition rate of $80$MHz, when
pumps have the same bandwidth.   Thus, as has been discussed before,
the conversion efficiency in the monochromatic pumps regime tends to
be significantly lower than for the pulsed-pumps regime.

\subsection{Pump bandwidth dependence} \label{bandw}

\begin{figure}[h!]
\begin{center}
\centering\includegraphics[width=5.5cm]{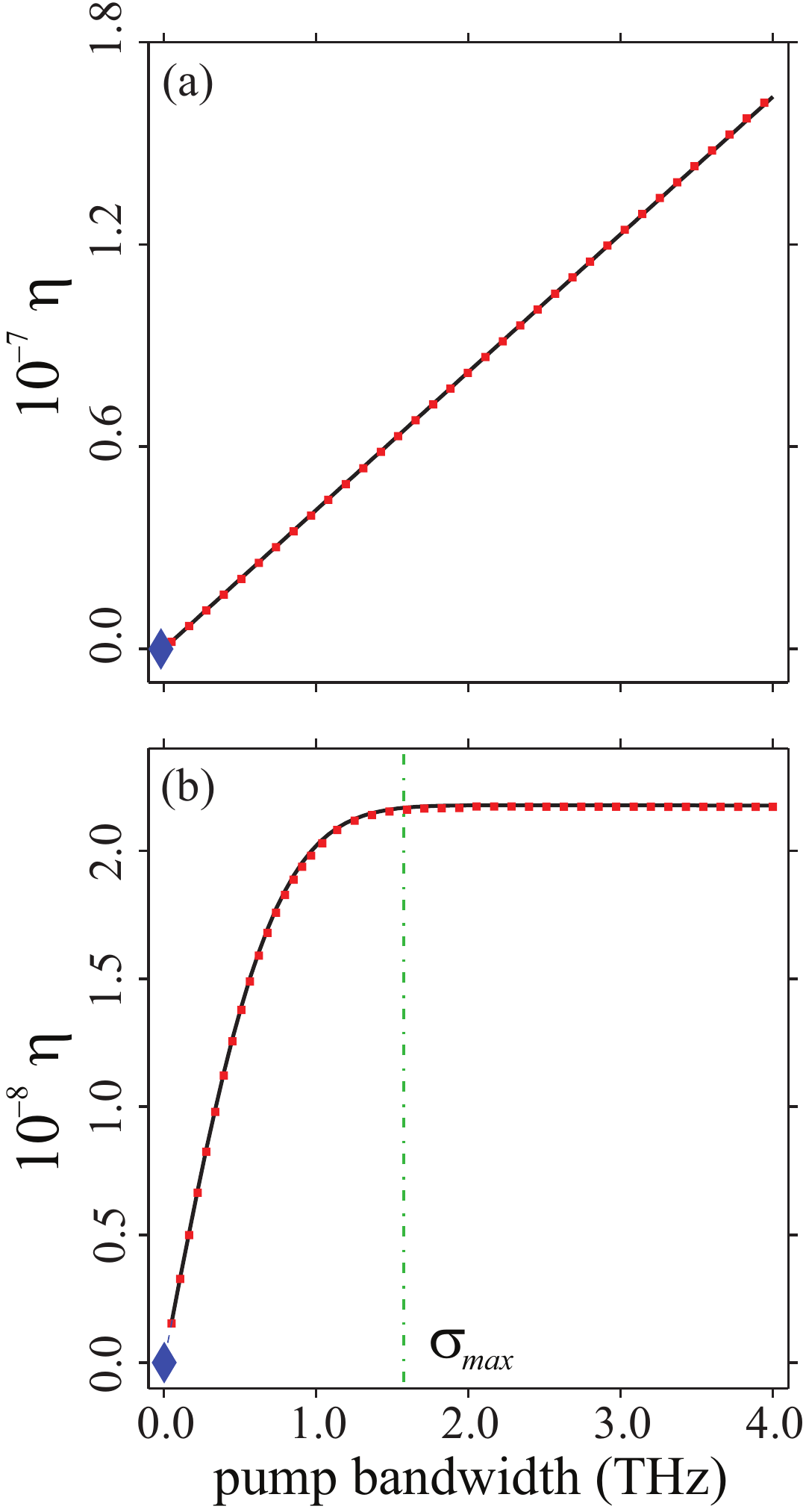}
\end{center}
\par
% Give a unique label
\caption{(Color online) Conversion efficiency as a function of pump bandwidth, for pulsed and monochromatic pump regimes (denoted by $\eta$ and $\eta_{cw}$, respectively), and for the following configurations: (a) degenerate pumps and (b) non-degenerate pumps. The red  squares are results obtained by numerical evaluation of Eq.~(\ref{etapul}), while the solid-black line corresponds to results obtained from analytical expressions described in section~\ref{anaexp}. The blue diamonds represent the conversion efficiency in the limit of monochromatic pumps.} \label{varwidth}
\end{figure}

In this section we turn our attention to the pump-bandwidth
dependence of the conversion efficiency (while maintaining the
energy per pulse, or alternatively the average power and the
repetition rate, in each of the two pump modes constant). We
consider the same source parameters as above.   Of course, as
$\sigma$ varies the pulse temporal duration varies, and consequently
the peak power varies too. We consider first the degenerate-pumps
configuration.   We evaluate the conversion efficiency from
Eqs.~(\ref{etapul}) and (\ref{etaDP}) for a  pump bandwidth $\sigma$
range $0.05-4.0$THz (which corresponds to a
Fourier-transform-limited temporal duration range $0.59-47.07$ps).
Numerical results (obtained from Eq.~(\ref{etapul})) as well as
analytical results (from Eq.~(\ref{etaDP})) are shown in
Fig.~\ref{varwidth}(a), exhibiting good agreement and, for small
$\sigma$, a linear dependence of the conversion efficiency on
$\sigma$.  Note that for large values of $\sigma$, there can be a
deviation from the linear trend apparent in Fig.~\ref{varwidth}(a).
Indeed, large values of $\sigma$ translate into greater spectral
width of the the signal and idler wave-packets, with the effect that
the linear approximation of the phasemismatch (upon which
Eq.~(\ref{etaDP}) is based) no longer suffices.  In the
monochromatic limit, evaluation of the conversion efficiency through
Eq.~(\ref{eficw}) predicts a value of
$\eta_{cw}=1.156\times10^{-11}$.  The blue diamond in
Fig.~\ref{varwidth}(a) indicates the conversion efficiency in this
limit;  it is graphically clear that the conversion efficiency
values for $\sigma \neq 0$ (calculated from Eq.~(\ref{etapul}))
approach the monochromatic-pumps limit (calculated from
Eq.~(\ref{eficw})).

We now consider the non-degenerate pumps case.  We vary $\sigma$ in
the range $0.05-4.0$THz and assume the same fiber length, average
pump power and repetition rate as for the degenerate-pumps case.
Fig.~\ref{varwidth}(b) shows numerical results indicated by red
squares superimposed with analytical results (from
Eq.~(\ref{etaNDP})) indicated by the solid-black line.  Let us note
that in this case, the flux dependence vs $\sigma$ is linear for
small $\sigma$, and exhibits a saturation effect around a specific
value of $\sigma$ to be referred to as $\sigma_{max}$.  In order to
understand this saturation effect, we note that while the pump
pulses remain temporally overlapped (as in the degenerate case
discussed above), the conversion rate has a linear dependence on
$\sigma$.  However, for non-degenerate pumps, the two pump modes
propagate at a different group velocities, and the resulting maximum
interaction length is given as in Eq.~(\ref{leff}), which for
$\sigma_1=\sigma_2 \equiv \sigma$ scales as $\sigma^{-1}$.   This
interaction length which decreases with $\sigma$ offsets the
conversion efficiency per unit fiber length which scales as
$\sigma$, leading to the saturation effect.  The bandwidth at which
saturation occurs, $\sigma_{max}$, is defined as the value of
$\sigma$ which makes the argument of the erf function in
Eq.~(\ref{etaNDP}) equal to $2$ (for which the erf function attains
$99.5\%$ of its maximum value), which leads to the following
expression

\begin{equation}
\label{sigeff} \sigma_{max}=\frac{4}{ L\left|k^{(1)}(\omega_1^o)-k^{(1)}(\omega_2^o)\right|}.
\end{equation}

In the specific case which we have modeled, $\sigma_{max}=1.58$THz,
which is indicated by a vertical dot-dashed line in
Fig.~\ref{varwidth}(b). As shown in the figure, the largest
considered pump bandwidth leads to a conversion efficiency of
$\eta=2.18\times10^{-8}$, which corresponds to $5.13\times10^7$
photon pairs emitted per second.  The corresponding conversion
efficiency obtained when pumps are in the monochromatic limit is
$\eta_{cw}=8.7\times10^{-12}$ and is indicated in the figure by the
blue diamond.  It is graphically clear that the conversion
efficiency values for $\sigma \neq 0$ (calculated from
Eq.~(\ref{etapul})) approach the monochromatic-pumps limit
(calculated from Eq.~(\ref{eficw})).  We note that if the spectral
separation between the two pump modes is not too great, temporal
broadening of the pump pulses due to second-order and higher-order
dispersion effects lead to some overlap between the pulses in the
two modes even for large values of $\sigma$.  This implies that
under these circumstances, the conversion efficiency does not reach
a strict plateau for $\sigma>\sigma_{max}$.

Let us note that the behavior of SFWM sources in terms of the
pump-bandwidth dependence of the conversion efficiency is different
from that observed for photon-pair sources based on SPDC.  In the
case of SPDC sources, the corresponding dependence is constant
within the phasematching bandwidth of the nonlinear crystal.  As in
the discussion related to pump-power dependence, SFWM sources
behave, in terms of the observed linear dependence of the conversion
efficiency on pump bandwidth, as a second-order nonlinear stimulated
nonlinear process, such as second harmonic generation, would. This
has an important consequence: because the spontaneous Raman
scattering flux, which tends to degrade the quality of the source,
exhibits a constant dependence on $\sigma$ \cite{lin07}, by
increasing $\sigma$ (which corresponds to shortening the temporal
duration of pump pulses), we can diminish the relative weight of
spontaneous Raman scattering as a fraction of the total emitted
light.

\subsection{Dependence on type/degree of spectral correlations}

We have shown in Refs.~\cite{garay07,garay08,garay08a} that it is
possible to engineer the spectral entanglement properties of photon
pairs generated by SFWM sources. Recent experimental
results~\cite{cohen09, halder09,soller10} confirm the ability of
engineered sources to generate, in particular, factorable photon
pairs.  In this context, it is natural to ask how the conversion
efficiency depends on the type of spectral correlations observed, a
question which represents the focus of this section.

In Ref.~\cite{garay07} we showed that the spectral entanglement
properties of the generated photon pairs can be controlled by the
pump frequency.    For a fiber with two zero-dispersion points
within the spectral region of interest, which can be the case of a
photonic crystal fiber (PCF),  the generated signal and idler
frequencies form a loop in the space of generated frequencies vs
pump frequency.   Each point around the loop corresponds to a
specific angle of orientation (covering all possible values between
$0$ and $2 \pi$ radians) of the phasematching function in the space
of generated frequencies $\{\omega_s,\omega_i\}$.  We will
illustrate our discussion considering a degenerate-pumps SFWM source
based on a PCF with a core radius of $r=0.5\mu$m and an air-filling
fraction of $f=0.6$, which exhibits zero dispersion points at
$0.6592 \mu$m and at $0.8595\mu$m (note that this is a different
fiber geometry from that assumed for the last three subsections).
For a pump wavelength $\lambda_p=0.75\mu$m, we obtained a value of
the non-linearity coefficient of $\gamma=337$km$^{-1}$W$^{-1}$,
through numerical integration of Eq.~(\ref{gamma}). We assume the
following choice of parameters: average pump power of $p=300\mu$W,
pump bandwidth of $\sigma=5$THz and fiber length of $L=1 $m. Let us
note that we have assumed a sufficiently large pump bandwidth, so
that the photon-pair correlations are determined by the
phasematching function rather than by the pump pulse bandwidth.

\begin{figure}[h!]
\begin{center}
\centering\includegraphics[width=6.5cm]{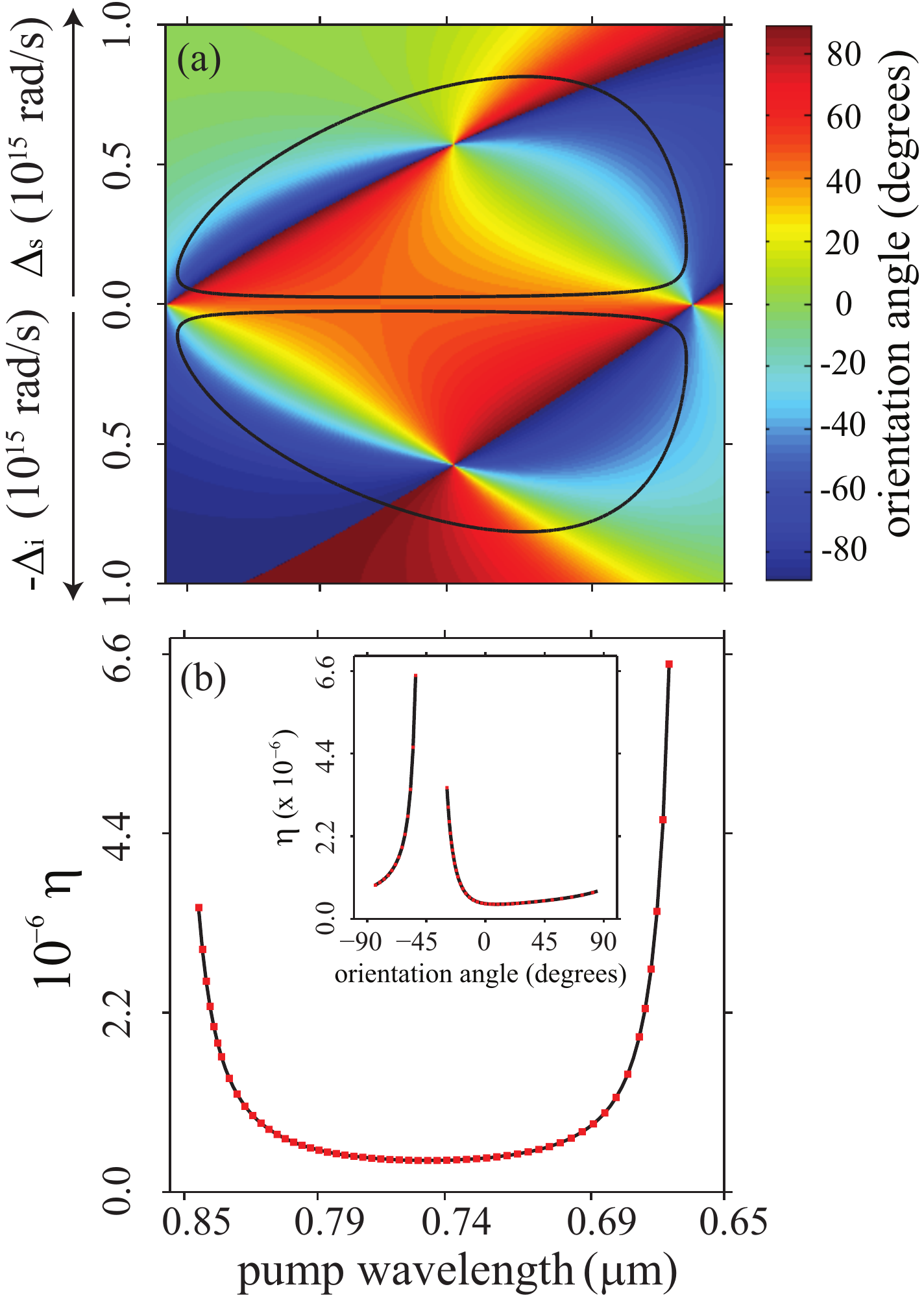}
\end{center}
\par
% Give a unique label
\caption{(Color online) (a) Solid-black curve: phase-matching
($\Delta k=0$) contour for SFWM in the degenerate pumps case.
Colored background: phase-matching orientation angle.  Note that the
pump frequency axis has been labeled with the corresponding
wavelength values. (b) Conversion efficiency as a function of the
pump frequency, varied within the range for which perfect
phase-matching occurs. The inset corresponds to the conversion
efficiency $\eta$, expressed as a function of the orientation
angle.} \label{varcorr}
\end{figure}

Figure~\ref{varcorr}(a) shows a plot of the $\Delta k=0$ contour
(the solid-black curve) in the $\{\omega_p,\Delta_{s,i}\}$ space,
where the generated frequencies are expressed as detunings from the
pump frequency $\Delta_{s,i}=\omega_{s,i}-\omega_{p}$. Likewise, in
this figure the phasematching orientation angle ($\theta_{si}$), in
the $\{\omega_s,\omega_i\}$ space, is represented by the colored
background, where it can be seen that this ranges from
$\theta_{si}=-90^{\circ}$ to $\theta_{si}=+90^{\circ}$, indicated in
blue and red respectively. Note that phasematching occurs within a
range of approximately $200$nm. Here, we will concentrate on the
`outer-branch' solutions (as opposed to the `inner-branch' solutions
which flank the pump frequency, represented by $\Delta_{s,i}=0$, at
a much smaller spectral separation). The outer branch is
sufficiently removed from the pump that contamination by photons
generated by spontaneous Raman scattering may be avoided.

%it can be prepared two-photon states in all degrees of spectral correlation, which is controlled by the pump frequency, when the pump-pulse bandwidth and fiber length are chosen such that the orientation of the JSA is determined by the phase-matching function\cite{garay07}. It is natural to expect that each engineered photon-pair source is characterized by its own spectral density, and therefore by a different emission rate. In this section, we are particularly interested in showing the dependence of the emitted flux by SFWM on the degree of spectral correlation of the two-photon state. To do this we consider that the photon-pair source is constructed in a PCF with a core radius $r=0.50\, \mu$m and a air-filling fraction $f=0.61$, which exhibits zero dispersion at $0.651\, \mu$m. In our simulations we assume that the nonlinear coefficient of PCF is almost constant within the considered range of pump frequency with a value $\gamma=160$ km$^{-1}$W$^{-1}$. Here, we restrict the study to a configuration with degenerate pumps, in which the pump power, pump bandwidth and fiber length are fixed parameters at $P_{av}=300\,\mu$W, $\sigma=7$ THz and $L=1$ m, respectively. Under these condition, for each pump frequency the orientation of the JSA corresponds to that of the phase-matching function.

We compute the conversion efficiency as a function of the pump
frequency, covering the following wavelength range:
$0.666-0.843\mu$m.  Fig.~\ref{varcorr}(b) shows the conversion
efficiency obtained numerically from Eq.~(\ref{etapul}) (red
squares) and the conversion efficiency obtained analytically from
Eq.~(\ref{etaDP}) (solid-black line). Because in general to each
pump frequency corresponds a given phasematching orientation angle
value, these results can be also plotted as a function of the
orientation angle, as shown in the inset of Fig.~\ref{varcorr}(b).
From these results, it is clear that anti-correlated two-photon
states, characterized by $\theta_{si}=-45^{\circ}$, lead to a larger
conversion efficiency as compared to other orientations in
$\{\omega_s,\omega_i\}$ space.  The physical reason for this greater
conversion efficiency is that for $\theta_{si}=-45^{\circ}$, the
phasematching function overlaps the pump envelope function over a
wider spectral range, leading to a greater generation bandwidth
which tends to enhance the value of the integral in
Eq.~(\ref{etapul}). This behavior is also consistent with
Eqs.~(\ref{etaNDP}) and (\ref{etaDP}), where if $k_s^{(1)}=
k_i^{(1)}$, which corresponds to $\theta_{si}=-45^{\circ}$, the
conversion efficiency diverges (within the linear approximation of
the phasemismatch) and increases markedly if higher-order terms are
taken into account.   Note that the agreement between numerical and
analytical results in Fig.~\ref{varcorr}(b) is excellent. While we
concentrated our discussion in this section on the degenerate-pumps
case, very similar conclusions apply for the non-degenerate case.

\section{Conclusions}

In this paper we have focused on the conversion efficiency of pump
photons into signal and idler photon pairs in the process of
co-polarized, spontaneous four-wave-mixing in single-mode optical
fibers.  We have derived expressions for the conversion efficiency,
defined as the signal-mode, single-photon flux divided by the pump
flux, as a function of all relevant experimental parameters.   Our
analysis includes on the one hand both the monochromatic and the
pulsed pump regimes, and on the other hand both the degenerate- and
non-degenerate-pumps configurations. These expressions are written
in terms of two-dimensional integrals, which for the case of pulsed
pumps we take to closed analytic form under certain approximations.
We present plots of the conversion efficiency as a function of
experimental parameters, including: fiber length, pump power and
pump bandwidth, computed through numerical integration of our
conversion efficiency expressions. We verify that the corresponding
conversion efficiency values computed from our expressions in closed
analytic form, for pulsed pumps, are in good agreement.   We find
that the behavior of the conversion efficiency with respect to pump
power and pump bandwidth is strikingly different from that observed
for spontaneous parametric downconversion.  In particular, the
linear dependence of the conversion efficiency on pump power
(compared to the constant dependence observed for SPDC) and the
linear dependence of the conversion efficiency on pump bandwidth in
the case of degenerate pumps (compared to the constant dependence
observed for SPDC) favor the design of bright photon pair sources.
We hope that this work will be useful for the design of fiber-based,
photon-pair sources, for the next generation of quantum-information
processing experiments.

\begin{acknowledgements}
This work was supported in part by CONACYT, Mexico,  by DGAPA, UNAM and by FONCICYT project 94142.
\end{acknowledgements}


\begin{thebibliography}{99}

\bibitem {zeilinger99} A. Zeilinger,  Rev. Mod. Phys. \textbf{71}, 288–297 (1999).      %1

\bibitem {kok07}  P. Kok,  W.J. Munro, K. Nemoto, T.C. Ralph, J.P. Dowling, and  G.J. Milburn,  Rev. Mod. Phys. \textbf{79}, 135–-174 (2007).  %2

\bibitem {burnham70}  D. C. Burnham and D. L. Weinberg,  Phys. Rev. Lett. \textbf{25}, 84, 135–-174 (1970).  %3

\bibitem {sharping01}  J. E. Sharping, M. Fiorentino, A. Coker, P. Kumar, and R. S. Windeler,  Opt. Lett. \textbf{26}, 1048--1050 (2001).   %4

\bibitem {Lee06} K. F. Lee, J. Chen, Ch. Liang, X. Li, P. L. Voss, and P. Kumar,   Opt. Lett. \textbf{31}, 1905--1907 (2006).%5

\bibitem{garay07} K. Garay-Palmett, H. J. McGuinness, Offir Cohen, J.S. Lundeen, R. Rangel-Rojo, M. G. Raymer, C.J. McKinstrie, S. Radic, A. B.
U'Ren, and I.A. Walmsley,   Opt. Express \textbf{15}, 14870--14886 (2007).  %6

\bibitem{Arbore} M. A. Arbore, M. M. Fejer, M. E. Fermann, A. Hariharan, A. Galvanauskas, and D. Harter,  Opt. Lett. \textbf{22}, 13--15 (1997). %7

\bibitem{Krischek10} R. Krischek, W. Wieczorek, A. Ozawa, N. Kiesel, P. Michelberger, T. Udem, and H. Weinfurter, Nat. Photon. \textbf{4}, 170–-173, (2010). %8

\bibitem{fulconis07} J. Fulconis, O. Alibart, W. J. Wadsworth, and J. G. Rarity,  New J. Phys. \textbf{9} 276 (2007). %9

\bibitem{chen05} J. Chen, X. Li, and P. Kumar,  Phys. Rev. A \textbf{72}, 033801 (2005). %10

\bibitem{alibart06} O. Alibart, J. Fulconis, G. K. L. Wong, S. G. Murdoch, W. J. Wadsworth, and J. G. Rarity,  New J. of Phys., \textbf{8}, 67 (2006).%11

\bibitem{brainis09} E. Brainis,  Phys. Rev. A \textbf{79}, 023840 (2009).%12

\bibitem {mandel}L. Mandel and E. Wolf, \textit{Optical Coherence and Quantum Optics}\/ (Cambridge University Press, 1995). %13

\bibitem{Rubin94} M. H. Rubin, D. N. Klyshko, Y. H. Shih, and A. V. Sergienko,  Phys. Rev. A \textbf{50}, 5122--5133 (1994). %14

\bibitem {agrawal2007}G. P. Agrawal, \textit{Nonlinear Fiber Optics, 4th Ed.} (Elsevier, 2007). %15

\bibitem{NLshift}  The nonlinear contribution to the phasemismatch can be shown to be given by $(\gamma_1+2\gamma_{21}-2\gamma_{s1}-2\gamma_{i1}) P_1+(\gamma_2+2\gamma_{12}-2\gamma_{s2}-2\gamma_{i2}) P_2$, where $\gamma_1$ and $\gamma_2$ result from self-phase modulation and the remaining $\gamma$ terms correspond to various cross-phase modulation contributions.   It may be shown that for the sources considered in this paper, the following represent valid approximations: $\gamma_1 \approx \gamma_{21} \approx \gamma_{s1} \approx \gamma_{i1}$ and  $\gamma_2 \approx \gamma_{12} \approx \gamma_{s2} \approx \gamma_{i2}$.  Under these approximations, we arrive at Eq.~(\ref{eq: delk}). %16

\bibitem{gammas} Note that while $\gamma_1$ and $\gamma_2$ are related to cross- and self-phase modulation effects, $\gamma$ governs the SFWM interaction. %17

\bibitem{tong04} L. Tong, J. Lou, and E. Mazur,  Opt. Express, \textbf{12}, 1025--1035 (2004). %18

\bibitem{foster04} M. A. Foster, K. D. Moll, and A. L. Gaeta,  Opt. Express, \textbf{12}, 2880--2887 (2004). %19

\bibitem{benson00} O. Benson, C. Santori, M. Pelton, and Y. Yamamoto, Phys. Rev. Lett. \textbf{84}, 2513  (2000). %20

\bibitem {garay08}K. Garay-Palmett, A. B. U'Ren, R. Rangel-Rojo, R. Evans, and S. Camacho-Lopez,  Phys. Rev. A \textbf{78}, 043827 (2008).  %21

\bibitem {rarity05}J. Rarity, J. Fulconis, J. Duligall, W. Wadsworth, and P. St. J. Russell,
Opt. Express \textbf{13}, 534--544 (2005). %22

\bibitem {fan05}J. Fan and A. Migdall,  Opt. Express \textbf{13}, 5777--5782 (2005).%23

\bibitem {cohen09} O. Cohen, J. S. Lundeen, B. J. Smith, G. Puentes, P. J. Mosley, and I. A. Walmsley, Phys. Rev. Lett. \textbf{102}, 123603 (2009). %24

\bibitem {Wong2005}G. K. L. Wong, A. Y. H. Chen, S. W. Ha, R. J. Kruhlak, S.
G. Murdoch, R. Leonhardt, J. D. Harvey and N. Y. Joly,  Opt. Express \textbf{13}, 8662--8670 (2005). %25

\bibitem{major06} A. Major, V. Barzda, P. A. E. Piunno, S. Musikhin, and U. J. Krull,  Opt. Express \textbf{14}, 5285-5294 (2006).%26

\bibitem{lin07}   Q. Lin, F. Yaman, and G. P. Agrawal,  Phys. Rev. A, \textbf{75}, 023803 (2007). %27

\bibitem{garay08a} K. Garay-Palmett, R. Rangel-Rojo, and A. B. U'Ren,  J. Mod. Opt. \textbf{55}, 3121-3131 (2008). %28

\bibitem {halder09} M. Halder, J. Fulconis, B. Cemlyn, A. Clark, C. Xiong, W. J. Wadsworth, and J. G. Rarity,  Opt. Express \textbf{17}, 4670--4676 (2009).%29

\bibitem{soller10}C. S\"{o}ller, B. Brecht, P.J. Mosley, L.Y. Zang, A. Podlipensky, N.Y. Joly, P.St.J. Russell, C. Silberhorn, Phys. Rev. A \textbf{81}, 031801(R) (2010). %30



\end{thebibliography}
\end{document}